\documentclass[aps,prm,showpacs,floatfix,twocolumn,byrevtex,superscriptaddress]{revtex4-2}
\usepackage{amsmath}
\usepackage{amssymb}
\usepackage{amstext}
\usepackage{amsopn}
\usepackage{amsfonts}
\usepackage{amsxtra}
\usepackage[english]{babel}
\usepackage{graphicx}
\usepackage{float}
\usepackage{bm}
\usepackage{multirow}
\usepackage{dcolumn}
\usepackage{color}
\usepackage{hyperref}

\newcommand{\NSNO}{Nd$_{1-x}$Sr$_{x}$NiO$_{2}$}
\newcommand{\NNO}{Nd$_{6}$Ni$_{5}$O$_{12}$}
\newcommand{\LN}{La$_{3}$Ni$_{2}$O$_{7}$}

\newcommand{\LNO}{La$_{3}$Ni$_{2}$O$_{7-\delta}$}

\newcommand{\LaNiO}{La$_{4}$Ni$_{3}$O$_{10}$}
\newcommand{\LAO}{LaAlO$_{3}$}

\begin{document}

\title{Growth and characterization of the La$_{3}$Ni$_{2}$O$_{7-\delta}$ thin films: dominant contribution of the $d_{x^{2}-y^{2}}$ orbital at ambient pressure}
\author{Yuecong Liu}
\author{Mengjun Ou}
\affiliation{National Laboratory of Solid State Microstructures and Department of Physics, Collaborative Innovation Center of Advanced Microstructures, Nanjing University, Nanjing 210093, China}
\author{Haifeng Chu}
\affiliation{College of Electronic Engineering, Nanjing Xiaozhuang University, Nanjing, 211171, People’s Republic
of China}
\author{Huan Yang}
\author{Qing Li}
\author{Ying-Jie Zhang}
\author{Hai-Hu~Wen}
\email{hhwen@nju.edu.cn}
\affiliation{National Laboratory of Solid State Microstructures and Department of Physics, Collaborative Innovation Center of Advanced Microstructures, Nanjing University, Nanjing 210093, China}

\date{\today}
%%%%%%%%%%%%%%%%%%%%%%%%%%%%%%%%%%%%
%
% Abstract
%

\begin{abstract}
By using the pulsed-laser-ablation technique, we have successfully grown the La$_{3}$Ni$_{2}$O$_{7-\delta}$ thin films with $c$-axis orientation perpendicular to the film surface. X-ray diffraction shows that the (00l) peaks can be well indexed to the La$_{3}$Ni$_{2}$O$_{7-\delta}$ phase. Resistive measurements show that the samples can be tuned from weak insulating to metallic behavior through adjusting the growth conditions. Surprisingly, all curves of $\rho-T$ in the temperature region of 2$\sim$300~K do not show the anomalies corresponding to either the spin density wave or the charge density wave orders as seen in bulk samples. Hall effect measurements show a linear field dependence with the dominant hole charge carriers, but the Hall coefficient $R_{H}=\rho_{xy}/H$ exhibits strong temperature dependence. The magnetoresistance above about 50~K is positive but very weak, indicating a weakened or absence of multiband effect. However, a negative magnetoresistance is observed at low temperatures, which shows the delocalization effect by magnetic field. Detailed analysis on the magnetoresistance suggests that the delocalization effect at low temperatures is due to the Kondo-like effect, rather than the Anderson weak localization. Our transport results suggest that, the electronic conduction is fulfilled by the $d_{x^{2}-y^{2}}$ orbital with holes as the dominant charge carriers, while the interaction through Hund's coupling with the localized $d_{z^{2}}$ orbital plays an important role in the charge dynamics.
\end{abstract}

%  71.55.Ak  Metals, semimetals, and alloys
%  72.15.Eb  Electrical and thermal conduction in crystalline
%  78.20.-e  Optical properties of bulk materials and thin films
%  78.30.-j  Infrared and Raman spectra

%\pacs{78.20.-e, 78.30.-j}

\maketitle

%%%%%%%%%%%%%%%%%%%%%%%%%%%%%%%%%%%%%%%%%%%%%%%%%%%%%%%%%%%%%%%%%%%%%%%%%%%%%%
%
\section{INTRODUCTION}
% Introduction
%
Since the discovery of cuprate superconductors, much efforts have been made in order to find other superconductors which are analogous to the cuprates. Ni is adjacent to Cu in the periodic table, and some of Ni-based compounds share many similarities with cuprate superconductors in terms of crystal structure and band structures. Thus, researchers have pointed out that high-temperature superconductivity may be expected in nickelates~\cite{anisimov1999electronic, lee2004infinite, poltavets2010bulk, zhang2017large}. However, many trial experiments along this line did not lead to success. Until 2019, superconductivity with a transition temperature ($T_{c}$) around 9$\sim$15~K was discovered in infinite-layer \NSNO\ thin films~\cite{li2019superconductivity} in which the Ni ions are supposed to have a valence state close to +1, leading to a similar outer-shell electronic state of $3d^{9}$ as in cuprates. Since then, researchers have found superconductivity in infinite-layer nickelate thin films with variety of rare earth elements, such as La, Pr, etc., and hole doping through alkaline-earth elements~\cite{osada2020superconducting, osada2021nickelate, zeng2022superconductivity}. Signature of superconductivity was also found in \NNO\ thin films in which the Ni ions have similar electron filling state as in \NSNO\ thin films~\cite{pan2022superconductivity}. As far as we know, superconductivity is still absent in bulk samples of infinite-layer nickelates~\cite{li2020absence, wang2020synthesis}. In thin films, the highest $T_{c}^{onset}$ of infinite-layer nickelates has reached 20 K at ambient pressure~\cite{sun2023evidence}, and around 30~K at a high pressure of 12.1~GPa~\cite{wang2022pressure} and $T_c$ rises with pressure without a saturation yet.

Very recently, signature of superconductivity with $T_{c}^{onset}$ around 80~K was found in bilayer nickelates \LN\ single crystals under high pressure~\cite{sun2023signatures}, which has attracted great attention in the community. Subsequently, superconducting state with zero resistance in \LN\ under high pressure is confirmed~\cite{zhang2024high}. This means that nickelates are likely to become the third type of unconventional high-temperature superconductive family after cuprates and iron-based superconductors. This discovery can also stimulate the  research on the pairing mechanism of unconventional high-temperature superconductivity. Unlike cuprates and infinite-layer nickelates, the Ni ions in \LN\ exhibit a $3d^{7.5}$ electron configuration. Theoretical calculations show that strong inter-layer coupling of $3d_{z^{2}}$ orbital forms the inter-layer $\sigma$-bonding and anti-bonding bands which locate below and above the Fermi level, respectively. It is supposed that the $3d_{z^{2}}$ bonding band crosses the Fermi level under high pressure, and the inter-layer coupling may play a crucial role in the pairing for superconductivity in La$_{3}$Ni$_{2}$O$_{7}$~\cite{sun2023signatures, lu2024interlayer}.

Through continuous efforts of studies, superconductivity has also been observed in polycrystalline La$_{3}$Ni$_{2}$O$_{7}$~\cite{wang2024pressure}, La$_{3-x}$Pr$_{x}$Ni$_{2}$O$_{7}$~\cite{wang2023observation} and trilayer nickelates La$_{4}$Ni$_{3}$O$_{10}$~\cite{zhu2024superconductivity, li2024signature, sakakibara2024theoretical, zhang2024superconductivity, li2024structural}. However, superconductivity of nickelates with these Ruddlesden-Popper (RP) structure all needs a high pressure. Recently,  the phase diagram of \LN\ under pressure has been modified to a triangular shape~\cite{li2024pressuredriven}. It is expected that the minimum pressure to induce superconductivity in RP nickelates could be reduced so that more in-depth research can be conducted. A feasible solution would be the preparation and research of high-quality thin films. In thin films, the biaxial strain provided by the substrate and the influence of the interface can change not only the lattice parameters of the sample, but also the oxygen octahedral torsion in the perovskite structures~\cite{borisevich2010suppression, aso2013atomic, moon2014effect, fowlie2017conductivity, fowlie2019thickness}. A recent research indicated that the phase in RP nickelate thin films can be tuned by the strain of the substrate~\cite{cui2023strain}.

 In this paper, we report the successful fabrication of a series of thin films with \LNO\ as the dominant phase by pulsed laser deposition (PLD). Through adjusting the growth conditions, weak insulating and metallic behavior were obtained in the \LNO\ thin films. Unlike the previous study that the \LN\ film with high crystalline exhibits insulating temperature dependence~\cite{li2020epitaxial}, our thin films show a clear metallic behavior. The X-ray diffraction (XRD) pattern and rocking curves of our \LNO\ films indicate that they have good crystallinity, which may be one of the reasons for the good conduction and metallic behavior. The $\rho-T$ curves of all the samples do not show any anomalies corresponding to either the spin density wave (SDW) or the charged density wave (CDW) as seen in bulk samples~\cite{wang2024pressure, zhang2024high, khasanov2024pressureinduced, kakoi2024multiband, liu2023evidence, hosoya2008pressure, wu2001magnetic, fukamachi2001studies, wu2001magnetic}. We have also measured the Hall resistance and magnetoresistance (MR) of a thin film showing metallic behavior in a wide temperature region. The transverse resistivity $\rho_{xy}$ shows linear field dependence with dominant hole charge carriers, but the Hall coefficient exhibits strong temperature dependence. The MR above 50~K is positive but very weak, and negative at low temperatures. Our analysis suggests that holes originated from the $d_{x^2-y^2}$ orbital are the dominant charge carriers.

%%%%%%%%%%%%%%%%%%%%%%%%%%%%%%%%%%%%%%%%%%%%%%%%%%%%%%%%%%%%%%%%%%%%%%%%%%%%%%
%
\section{EXPERIMENTAL DETAILS}

 The \LNO\ thin films were grown on single-crystal (00$l$)-oriented \LAO\ substrates by the technique of pulsed-laser deposition, with substrate temperature at $800\rm{^\circ C}$ and laser frequency of 2~Hz for all cases. Prior to the deposition, a polycrystalline \LNO\ target was prepared by the sol-gel method~\cite{poltavets2006oxygen}. During the growth, the laser spot area of 4.68~$\rm{mm^2}$ was fixed, and two laser fluence of 1.52~$\rm{J/cm^2}$ (labelled as SL) and 2.07~$\rm{J/cm^2}$ (labelled as SH) were applied. And the oxygen partial pressure during the growth was 20$\sim$35~Pa. After growth, the samples were cooled down at a cooling rate of $8\rm{^\circ C/min}$ to $450\rm{^\circ C}$ under oxygen partial pressure $P_{O_{2}}$=5$\sim$8$\times10^4$~Pa and post-annealed in-situ at $450\rm{^\circ C}$ for about 1$\sim$48~h, followed by the same cooling procedure to room temperature. Then they were taken out for the subsequent measurements. XRD and X-ray reflectivity (XRR) measurements were performed by using a Bruker D8 Advanced diffractometer with Cu $\rm{K_{\alpha1}}$ radiation. The reciprocal space map (RSM) is measured by a Bruker D8 discover diffractometer. The morphology characterization obtained by a scanning electron microscope (SEM) (Fig.~\ref{Figure2}(a)) was conducted by using the secondary electron imaging at an accelerating voltage of 10~kV. The size of the crystalline and surface roughness were obtained by an atomic force microscopy (AFM). The surface diffraction pattern was obtained by an in-situ high-pressure reflection high-energy electron diffraction (HP-RHEED) with a voltage of 30~kV after growth. Transport properties were measured by a physical property measurement system (PPMS, Quantum Design) with a magnetic field up to 9~T.

%The \LO\ and \NOH\ powder with amounts of precisely chemical ratio was dissolved in concentrated \HNO\ solution, Then add an appropriate amount of citric acid and ethylene glycol to the solution. The liquid was baked at $300^\circ C$ for about 6h to get the gel,
% Experiment
%

%%%%%%%%%%%%%%%%%%%%%%%%%%%%%%%%%%
%%%%%%%%%%%%%%%%%%%%%%%%%%%%%%%%%%%%%%%%%%%%%%%%%%%%%%%%%%%%%%%%%%%%%%%%%%%%%%
%
\section{RESULTS AND DISCUSSION}
\subsection{Structural characterizations}
% Experiment
%

%%%%%%%%%%%%%%%%%%%%%%%%%%%%%%%%%%
% Figure 1
\begin{figure}[tb]
\includegraphics[width=\columnwidth]{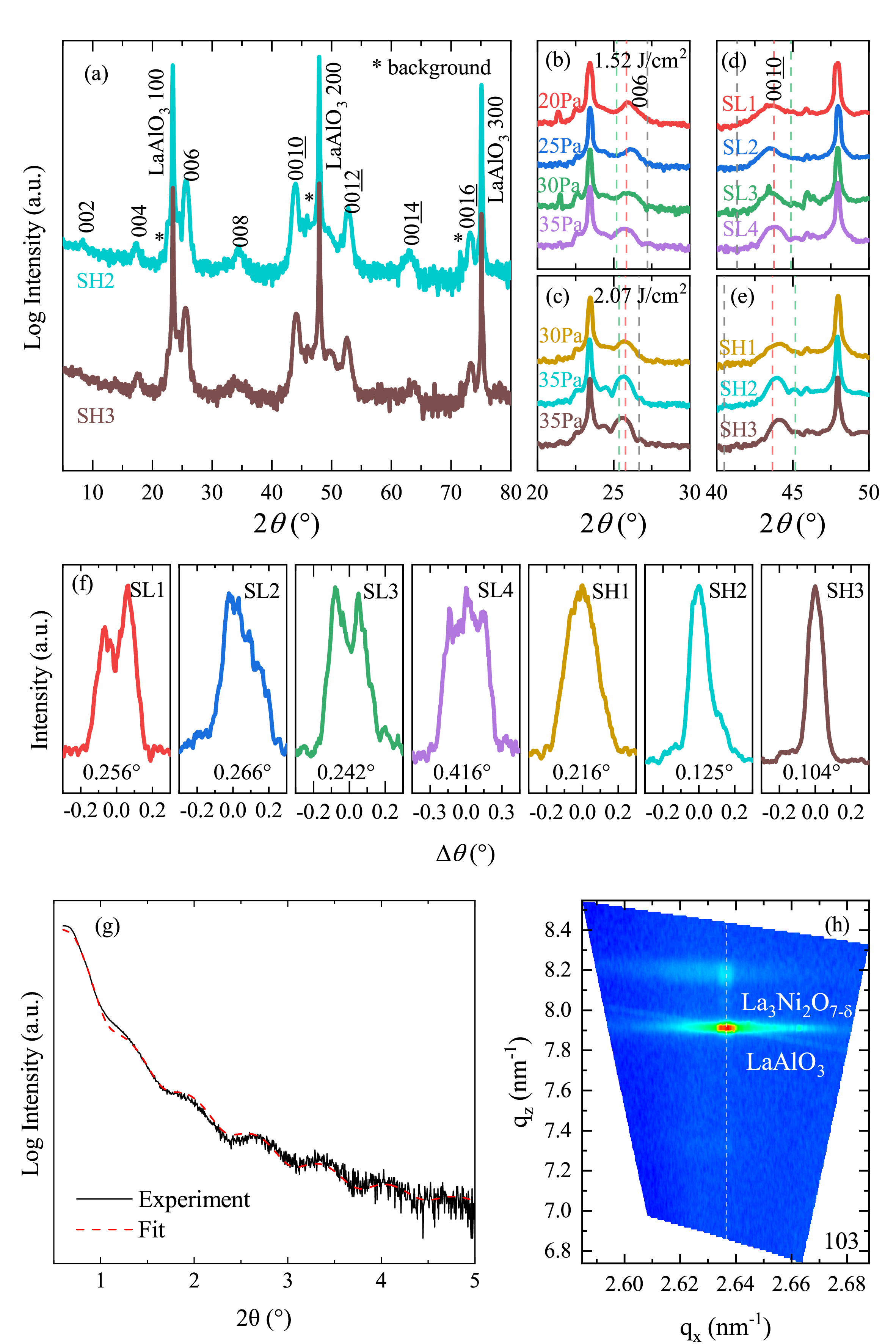}
\caption{Structural characterizations of the as-grown \LNO\ films. (a) Wide-range XRD 2$\theta$-$\omega$ scans of sample SH2 and SH3. (b)-(e) Enlarged XRD 2$\theta$-$\omega$ scans of \LNO\ thin films grown with different conditions around the (006) peak (b and c) and (0010) peak (d and e). The oxygen partial pressure is marked in (b) and (c) for each sample during growth. The dashed lines in grey, light red, and light green represent the peak positions of $\rm{La_2NiO_4}$, $\rm{La_3Ni_2O_{7-\delta}}$, and \LaNiO\ phases, respectively. (f) Rocking curves for the (006) peaks of the \LNO\ films, the full-width-at-half-maximum of the peaks are given by the text at the bottom of the figure. (g) X-ray reflectivity of sample SH3. The red dashed line shows the fit with film thickness $\sim$ 12.056nm. (h) Reciprocal space map of sample SH3.}
\label{Figure1}
\end{figure}
 Fig.~\ref{Figure1}(a) shows the wide-range XRD 2$\theta$-$\omega$ scans for the metallic \LNO\ thin films (sample SH2 and SH3). All the (00$l$) ($l$ is even) peaks of \LNO\ appeared without any obvious impurity peaks, suggesting the $c$-axis orientation of the films without any visible impurity phases. Fig.~\ref{Figure1}(b)-(e) show the enlarged XRD 2$\theta$-$\omega$ scans of as-grown films around the \LNO\ (006) peak (b and c) and (0010) peak (d and e). The dashed lines in grey, light red, and light green represent the peak positions of $\rm{La_2NiO_4}$, $\rm{La_3Ni_2O_{7-\delta}}$, and \LaNiO\ phases, respectively. The XRD 2$\theta$-$\omega$ measurements indicate that all of the thin films display the dominant \LNO\ phase.  Samples in Fig.~\ref{Figure1}(b) and (d) were grown with lower laser energy (laser fluence 1.52~$\rm{J/cm^2}$), and samples in Fig.~\ref{Figure1}(c) and (e) were grown with higher laser energy (laser fluence 2.07~$\rm{J/cm^2}$). The remarks in Fig.~\ref{Figure1}(b) and (c) mark the oxygen partial pressure for the growth of each sample. Sample SL4 and SH3 were post-annealed under oxygen partial pressure $P_{O_{2}}$ = $8\times10^4$~Pa at $450\rm{^\circ C}$ for 24 and 48~h, respectively. Other samples were post-annealed under oxygen partial pressure $P_{O_{2}}$ = $5\times10^4$~Pa at $450\rm{^\circ C}$ for 1~h. As shown in Fig.~\ref{Figure1}(f), the films grown with higher laser energy have better crystallinity than those with lower laser energy. Fig.~\ref{Figure1}(g) shows the X-ray reflectivity of sample SH3 with corresponding fit shown in a red dashed line, which yields a film thickness of about 12.056 nm. The thicknesses of samples SH1 and SH2 are around 20 nm. Other samples have also thickness around 12nm. Fig.~\ref{Figure1}(h) shows the reciprocal space map of sample SH3 around the (103) diffraction peak of the \LAO\ substrate. The dashed line indicates that the \LNO\ thin film has a in-plane constrain by the substrate. The diffraction peak broadening on the left-hand side of the \LNO\ thin film is slightly stronger than that on the right-hand side, which may be due to a slight relaxation in the surface layer. The $a$-axis constant is calculated to be between 3.792 \r A and 3.807 \r A, smaller than that of the bulk samples. And the $c$-axis constant is around 20.722 \r A, slightly larger than that of the bulk sample due to the in-plain compressive strain.

%%%%%%%%%%%%%%%%%%%%%%%%%%%%%%%%%%%%%%%%%%%%%%%%%%%%%%%%%%%%%%%%%%%%%%%%%%%%%%

%

%%%%%%%%%%%%%%%%%%%%%%%%%%%%%%%%%%
% Figure 2
\begin{figure}[tb]
\includegraphics[width=\columnwidth]{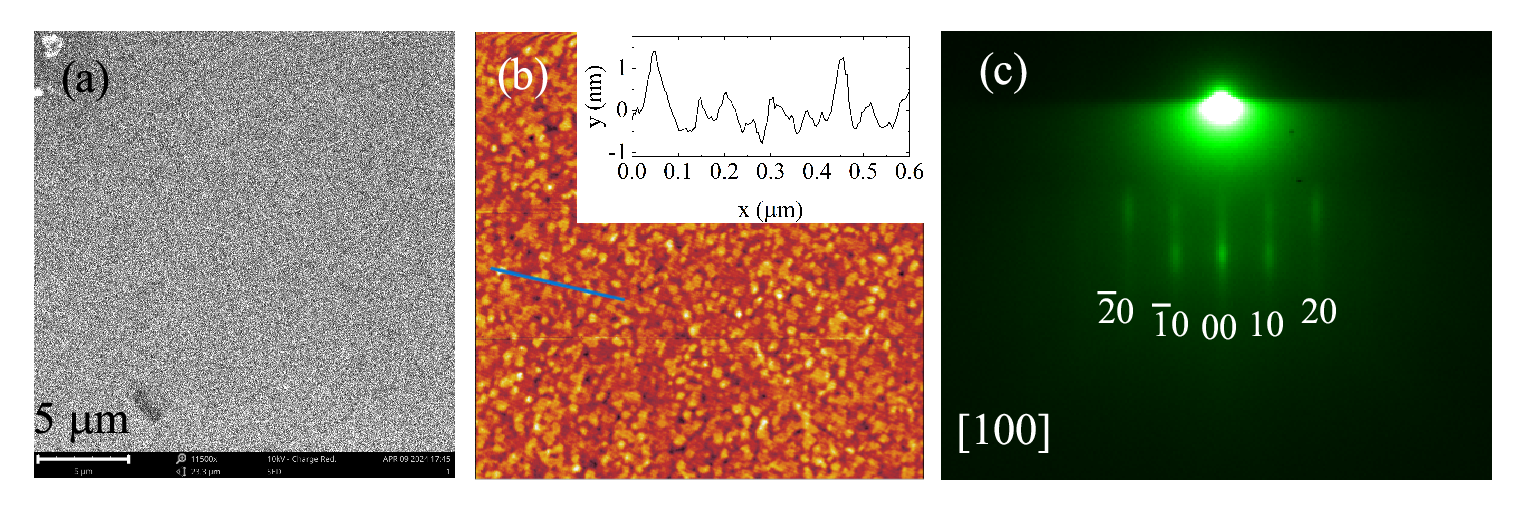}
\caption{Morphological and chemical composition distribution characterizations of sample SH3. (a) Secondary electron SEM image. (b) AFM image of a $2\rm{\mu m}\times2\rm{\mu m}$ area. The inset shows the corresponding line cut along the blue line. (c) RHEED pattern of sample SH3. }
\label{Figure2}
\end{figure}
Fig.~\ref{Figure2}(a) shows the scanning electron microscopy secondary electron image of sample SH3. The surface of the film is relatively flat. Fig.~\ref{Figure2}(b) shows the atomic force microscopy topography of sample SH3. The diameter of the grains is about 50$\sim$100~nm as shown in the inset of Fig.~\ref{Figure2}(b), and the average roughness of the surface is about 0.32~nm. Fig.~\ref{Figure2}(c) shows the RHEED pattern of sample SH3. The clear diffraction streaks at high oxygen partial pressure of 35~Pa indicate a flat surface of the film.

\subsection{Transport measurements}
%%%%%%%%%%%%%%%%%%%%
% Figure 3
\begin{figure}[tb]
\includegraphics[width=\columnwidth]{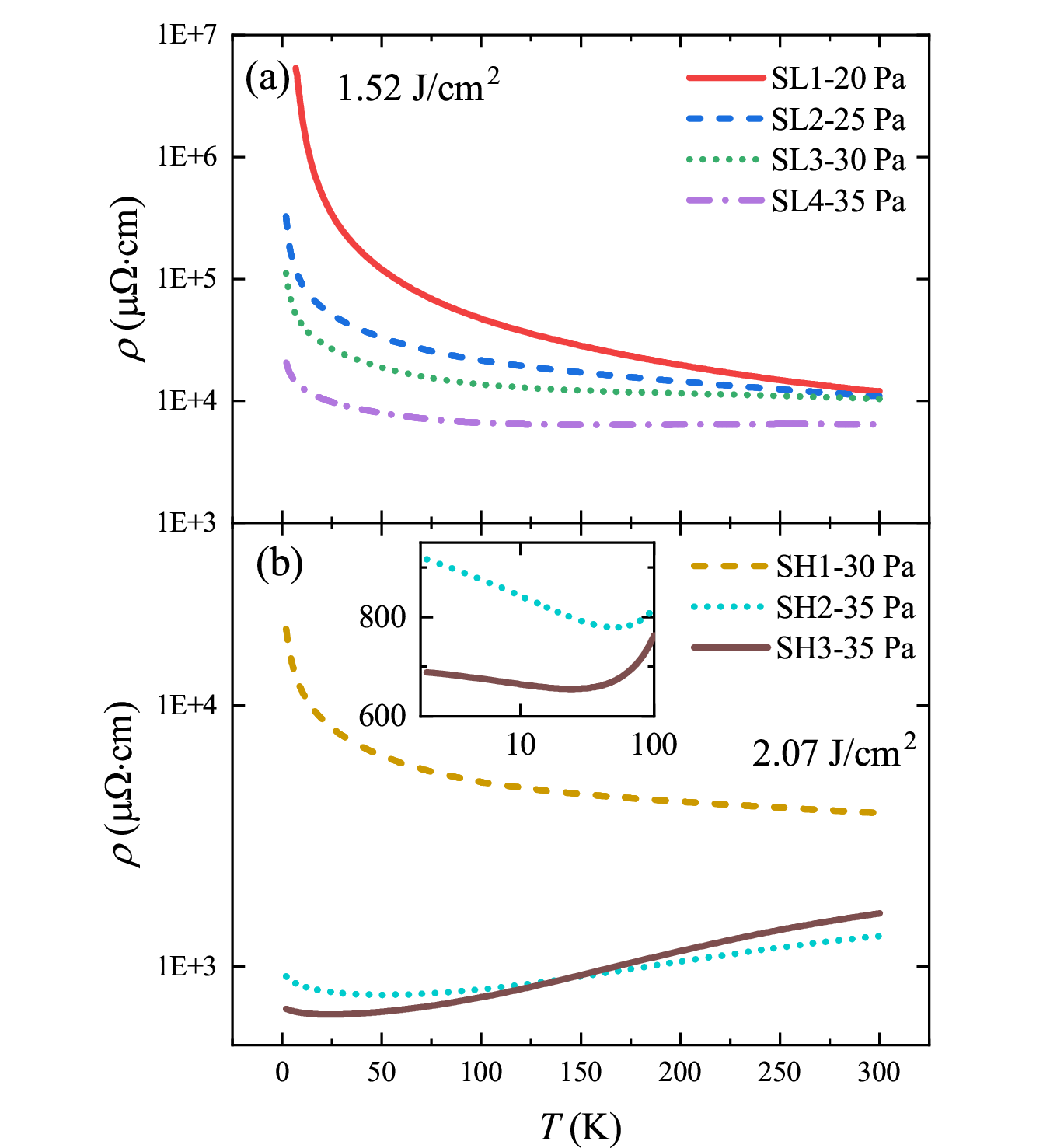}
\caption{Temperature-dependent resistivity of samples grown with lower (a) and higher laser energy (b). The inset of (b) shows an enlarged view of low temperatures in semi-logarithmic plot.}
\label{Figure3}
\end{figure}
Fig.~\ref{Figure3}(a) and (b) show the $\rho-T$ curves of \LNO\ thin films grown with lower and higher laser energy, respectively. It is clear that the resistivity of the samples decreases with the increasing oxygen partial pressure for both lower and higher deposition energy. However, as shown in Fig.~\ref{Figure3}(a), samples synthesized with lower laser energy are weak insulators for all deposition oxygen partial pressures and annealing conditions. We also tried growth with higher growth partial pressures of oxygen ($\sim$40~Pa) and with lower laser fluence. But the dominant phase of the thin film becomes \LaNiO\ instead of the metallic La$_{3}$Ni$_{2}$O$_{7-\delta}$. This may be the reason for the poor crystallinity of the La$_{3}$Ni$_{2}$O$_{7-\delta}$ in the these samples, as shown in Fig.~\ref{Figure1}(f). As shown in Fig.~\ref{Figure3}(b), the samples synthesized with higher laser energy are insulating when the growth oxygen partial pressure is less than 30~Pa. With the oxygen partial pressure raised to 35~Pa, the \LNO\ thin films demonstrate a clear metal-to-insulator transition as temperature decreases. Through long-time annealing at high oxygen partial pressure, the metal-to-insulator transition temperature can be lowered from 51.3~K (sample SH2) to 24.3~K (sample SH3). This may suggests that the weak insulating behavior of \LNO\ is related to oxygen vacancies, consistent with previous studies~\cite{kobayashi1996transport, ling2000neutron, zhang1994synthesis, taniguchi1995transport}. It is noteworthy that all the $\rho-T$ curves in the whole measurement temperature region do not show the DW-like or SDW-like anomalies as seen in bulk samples. The possible transition may be suppressed by the substrate strain, as that in the bulks is suppressed by physical pressure~\cite{zhang2024high, wang2024pressure, khasanov2024pressureinduced}. The inset of Fig.~\ref{Figure3}(b) shows the resistivity behavior of sample SH2 and SH3 at low temperatures on semi-log plots. A clear logarithmic increase of the resistivity with decreasing temperature is observed. This may correlate with Kondo effect or weak localization in a Quasi-two-dimensional (2D) system, as will be addressed below.

We also performed some ex-situ oxygen annealing on these samples. We annealed these samples in an atmosphere of oxygen using a thermal gravimetric analyzer. However, the samples became more insulating. This may be due to the heat-treatments which worsens the film crystallinity. We then tried to process the insulating samples at a high oxygen pressure of 8 MPa, but even after annealing for a long time ($\sim$60 h) at such a high oxygen pressure, they still could not become metallic, only a structural phase transition appeared at around 250 K. Unlike the bulk samples, where oxygen treatment can change the samples from insulating to metallic, for the film sample, it depends more on the growth process rather than on the oxygen treatment concerning whether it is metallic or insulating . Compared with the rocking curves in Fig.~\ref{Figure1}(f), the full-width-at-half-maximum on the \LNO\ (006) peaks of samples grown at lower laser energy are generally wider than those grown at higher energy. And the rocking curve peaks of metallic samples are much sharper than that of the weak insulating samples, indicating the crystallinity might be important to gain metallic \LNO\ thin films.

%%%%%%%%%%%%%%%%%%%%
% Figure 4
\begin{figure}[tb]
\includegraphics[width=\columnwidth]{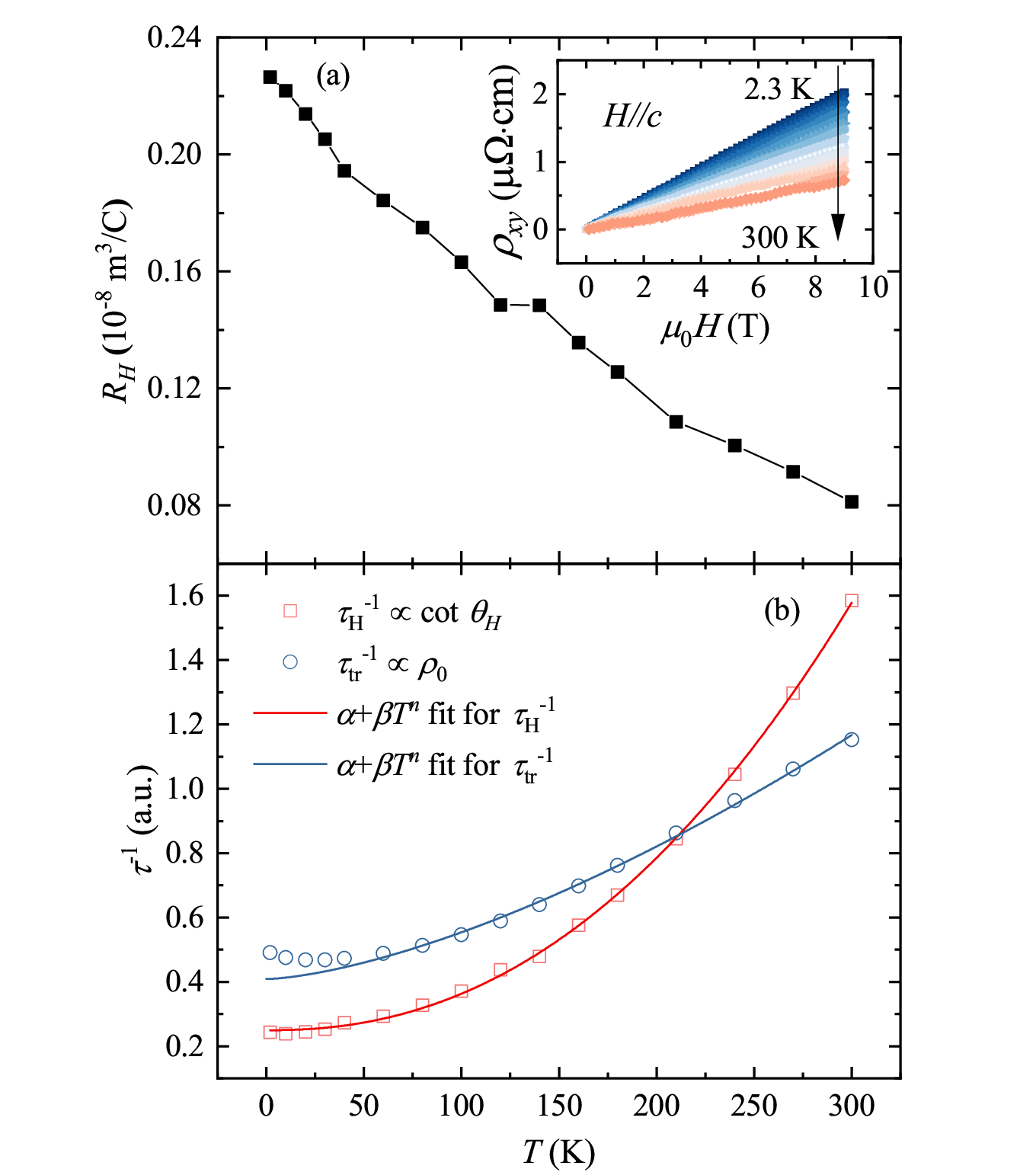}
\caption{(a) Temperature dependence of the Hall coefficient $R_{H}$ for sample SH3. The inset shows the magnetic field dependence of $\rho_{xy}$ at different temperatures. (b) Temperature dependence of $\tau_{H}^{-1}$ and $\tau_{tr}^{-1}$. The solid lines show the fit by $\alpha+\beta T^n$ with $n=2.24\pm0.03$ for $\tau_{H}^{-1}$ and $n=1.50\pm0.08$ for $\tau_{tr}^{-1}$.}
\label{Figure4}
\end{figure}
As shown in the inset of Fig.~\ref{Figure4}(a), the raw data of the transverse resistivity $\rho_{xy}$ shows good linear field dependence with a positive slope at all temperatures, indicating that the hole-like charge carriers dominate the electric conduction. The band structure calculation shows that the hole pockets are mainly contributed by $d_{x^{2}-y^{2}}$ orbital~\cite{sun2023signatures, yang2024orbital, luo2023bilayer, luo2023hightc}. These results suggest holes originating from $d_{x^{2}-y^{2}}$ orbital are the dominant charge carriers.
Judged from the data in Fig.~\ref{Figure4}(a), the value of Hall coefficient $R_{H}=\rho_{xy}/\mu_0H$ is comparable to that of the single crystal at pressures below 5~GPa~\cite{zhou2024evidence}. It is known that for a normal metal with Fermi liquid feature, the Hall coefficient should be a constant versus temperature. However, the $R_{H}$ of the \LNO\ films exhibits strong temperature dependence that decreases with the increasing temperature, this may be related to multiband effect or correlation effect.
We did the calculation of the Hall angle that $\rm{cot}~\theta_H=\rho_{xx}/\rho_{xy}\equiv1/(\omega_c\tau_H)$, which measures only the scattering rate $1/\tau$ and the information about the charge-carrier density is naturally separated away. Here the $\tau_H$ is the scattering time $\tau$ determined from Hall angle, $\omega_c$ is the circling frequency. The temperature dependence of $\tau_{H}^{-1}\propto\rm{cot}~\theta_H$ is shown in Fig.~\ref{Figure4}(b). And it can be well fitted by the expression of $\alpha+\beta T^n$ with $n=2.24\pm0.03$. The temperature dependence of $1/\tau_{tr}$ calculated from resistivity is also shown in Fig.~\ref{Figure4}(b). $1/\tau_{tr}$ exhibits a complex structure, which can reflect the combined results of the true scattering rate $1/\tau_{H}$ and the temperature dependent density of states (DOS). We also try to fit the metallic part of $1/\tau_{tr}\propto\rho_0$ with $\alpha+\beta T^n$, and a good fitting result with $n=1.50\pm0.08$ is obtained. The discrepancy between $1/\tau_{H}$ and $1/\tau_{tr}$ manifests that the effective DOS may have a strong temperature dependence that is depleted in lowering temperature, which is a common feature in correlated materials. And this might be part of the reason that the Hall coefficient $R_H$ exhibits strong temperature dependence.

%%%%%%%%%%%%%%%%%%%%
% Figure 5
\begin{figure}[tb]
\includegraphics[width=\columnwidth]{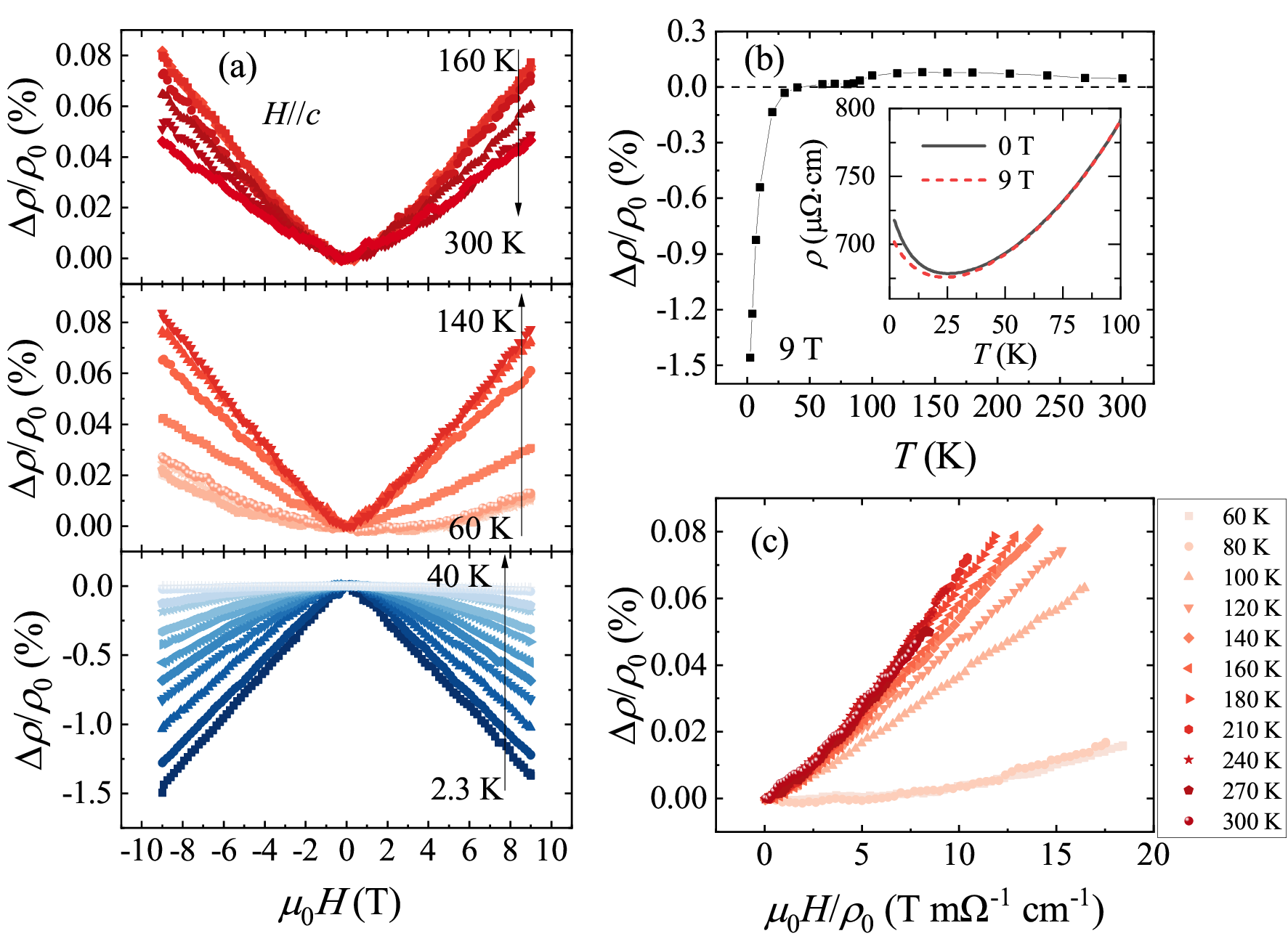}
\caption{(a) Field dependence of MR $\Delta\rho/\rho_{0}$ for sample SH3 at different temperatures. (b) Temperature dependence of the magnetoresistance at 9~T. The inset shows the $\rho-T$ curve at 0~T and 9~T. (c) The positive MR plotted by following Kohler's law at different temperatures.}
\label{Figure5}
\end{figure}
To further investigate the electronic scattering and the information of the Fermi surface, we systematically measured the magnetoresistance of the metallic \LNO\ thin film. The magnetoresistivity is defined as $\Delta\rho=\rho(H)-\rho_0$, where $\rho(H)$ is the longitudinal resistivity in a magnetic field $H$ and $\rho_0$ is that at a zero-field. As shown in Fig.~\ref{Figure5}(a), the magnetoresistance $\Delta\rho/\rho_0$ above 50~K is positive but very weak (less than 0.1$\%$), indicating a weakened or absence of multiband effect. Thus, we think that the strong temperature dependence of the Hall coefficient should be partially contributed by the multiband effect which becomes weaker in high temperature region (due to temperature enhanced interband scattering), and partially due to the correlation effect, because the effective DOS at Fermi energy is temperature-dependent and becomes depleted with decreasing temperature. The negative magnetoresistance at temperatures below 40~K in Fig.~\ref{Figure5}(a) indicates the delocalization effect in a magnetic field, and it decreases with increasing temperature. The MR above 60~K is positive, and with the increasing temperature, it increases in the temperature region of 60$\sim$140~K and decreases slightly at temperatures above 160~K. Fig.~\ref{Figure5}(b) shows the temperature dependence of the magnetoresistance at 9~T. It can be observed that the negative magnetoresistance at low temperatures exhibits a rapid decay with increasing temperature, while the positive magnetoresistance at high temperatures is notably weak and exhibits minimal variation with temperature. The inset of Fig.~\ref{Figure5}(b) shows the $\rho-T$ curve at 0~T and 9~T. The magnetic field exerts a slight suppression on the upturn of resistivity at low temperatures in a magnetic field of 9~T. According to Kohler's law~\cite{ziman2001electrons}, if there is only one isotropic scattering time $\tau$ in the transport properties of a system, the magnetoresistance $\Delta\rho/\rho_0=F(H\tau)=f(H/\rho_0)$, where $F$ and $f$ represent some unknown functions (assuming that the scattering rate $1/\tau(T)\propto\rho_0(T)$). Fig.~\ref{Figure5}(c) shows the positive MR plotted by following Kohler's law, One can see that Kohler's law is obeyed at temperatures above about 180~K, disobeyed at temperatures below 160~K. As mentioned before, the variation of DOS with temperature due to correlation effect can make the basic requirement of Kohler's law that $1/\tau(T)\propto\rho_0(T)$ invalid, which may cause the failure of Kohler's law for describing the magnetoresistance.

\subsection{Analysis for the resistivity and negative magnetoresistance at low temperatures}

%%%%%%%%%%%%%%%%%%%%
% Figure 6
\begin{figure}[tb]
\includegraphics[width=\columnwidth]{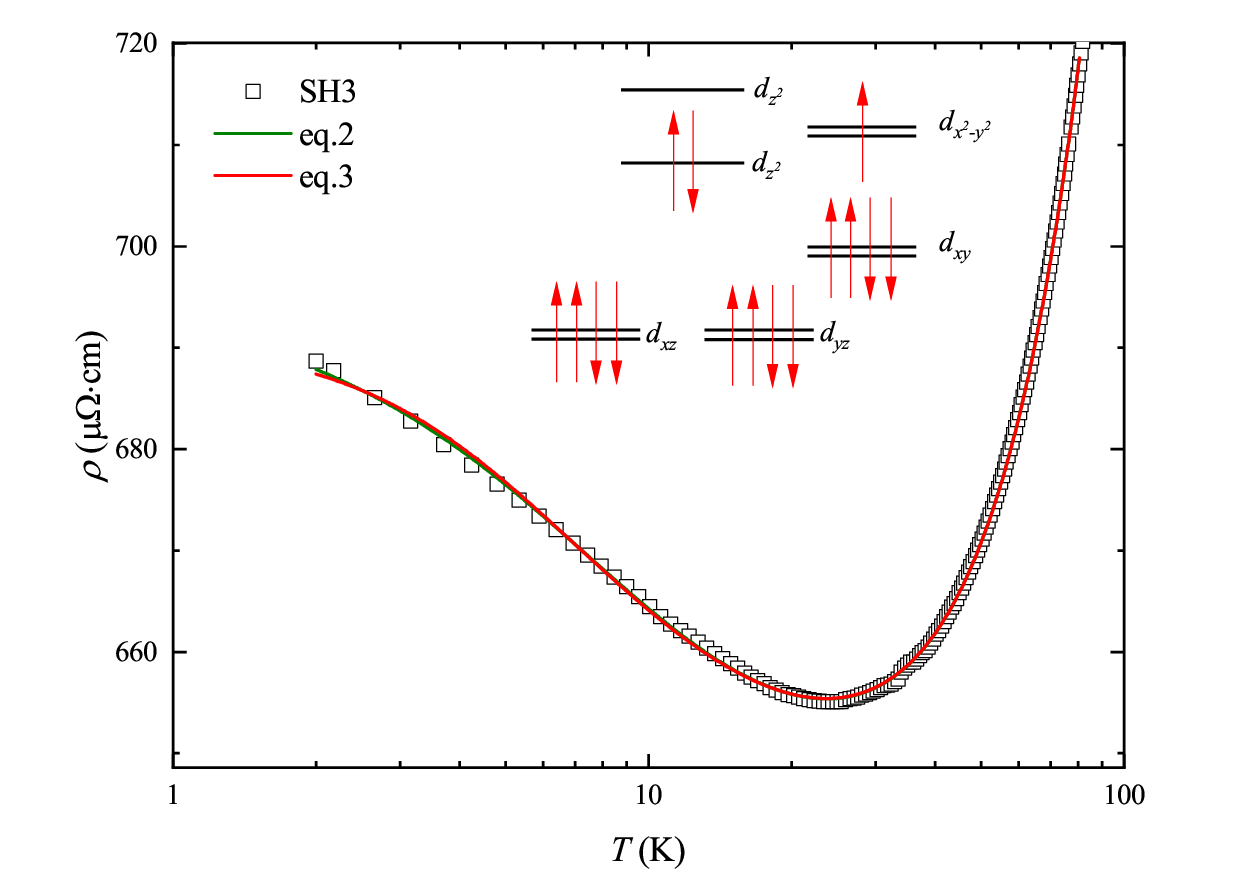}
\caption{$\rho-T$ curve and Kondo fit (eq.~\ref{KondoRk} and eq.~\ref{KondoRkb}) at low temperatures for sample SH3. The inset shows a schematic electronic configuration of \LNO.}
\label{Figure6}
\end{figure}

Now we turn to the negative magnetoresistance at low temperatures. It is found that this negative magnetoresistance is closely linked with the upturn of resistivity at low temperatures. The negative MR or the upturn of low-T resistivity may have two different origins: (1) Kondo effect or (2) Anderson localization in 2D systems. We first assume that the cause of the logarithmic increase in resistivity and negative MR at low temperatures is due to the Kondo effect, thus the $\rho-T$ curve at low temperatures can be fitted by

% Eq.1
%
\begin{equation}
\begin{aligned}
\rho(T) =\rho_{0}+aT^n+\rho_{K}(T)
\label{KondoR}
\end{aligned}
\end{equation}
Where $\rho_0$ is the residual resistivity, $\rho_{K}(T)$ is the Kondo scattering term that contributes to the logarithmic rise behavior at low temperatures, and it can be expressed by Hamann formula \cite{hamann1967new}

% Eq.2
%
\begin{equation}
\begin{aligned}
\rho_{K}(T) =\rho_{K_{0}}\bigg[1-\frac{ln(\frac{T}{T_{K}})} {\sqrt{ln^{2}(\frac{T}{T_{K}})+S(S+1)\pi^{2}}}\bigg]
\label{KondoRk}
\end{aligned}
\end{equation}
Where $\rho_{K_{0}}$ is a temperature-independent constant, $T_{K}$ is the Kondo temperature, $S$ is the spin of the magnetic impurities. In order to improve the reliability of fitting results, we first fit the $\rho-T$ curve in the temperature region of 50$\sim$80~K with $\rho=\rho_0+aT^n$ (since the negative magnetoresistance disappears at above 50 K and the temperature region for the Kondo fit is chosen for temperatures between 2 to 80 K), then put the value of the resulting $n\approx 2.217$ into the Kondo fitting formula and make sure that the values of $\rho_{0}$ and $a$ obtained from the Kondo fit are close to those obtained from the $\rho=\rho_0+aT^n$ fit for 50$\sim$80~K. The curve fitted by eq.~\ref{KondoRk} of the low-temperature resistivity is shown as the green solid line in Fig.~\ref{Figure6}, with $T_K\approx 8.14$~K and $S\approx 0.23$.

We also use an empirical formula for the resistivity to fit the $\rho-T$ curve~\cite{goldhaber1998kondo, lee2011electrolyte}

% Eq.3
%
\begin{equation}
\begin{aligned}
\rho_{K}(T) =\rho_{K_{0}}\bigg[\frac{T_{K}^{2}} {(2^{1/S}-1)T^{2}+T_{K}^{2}}\bigg]^{S}.
\label{KondoRkb}
\end{aligned}
\end{equation}
As shown by the red solid line in Fig.~\ref{Figure6}, it gives $T_K\approx 9.66$~K and $S\approx 0.45$, being close to the fitting results by eq.~\ref{KondoRk}. Since it is a fit with multiple fitting parameters, the final fitting result may subject to some errors, but the Kondo-like scattering picture seems to be substantiated by the general behavior of the integrated data system.

%%%%%%%%%%%%%%%%%%%%
% Figure 7
\begin{figure}[tb]
\includegraphics[width=\columnwidth]{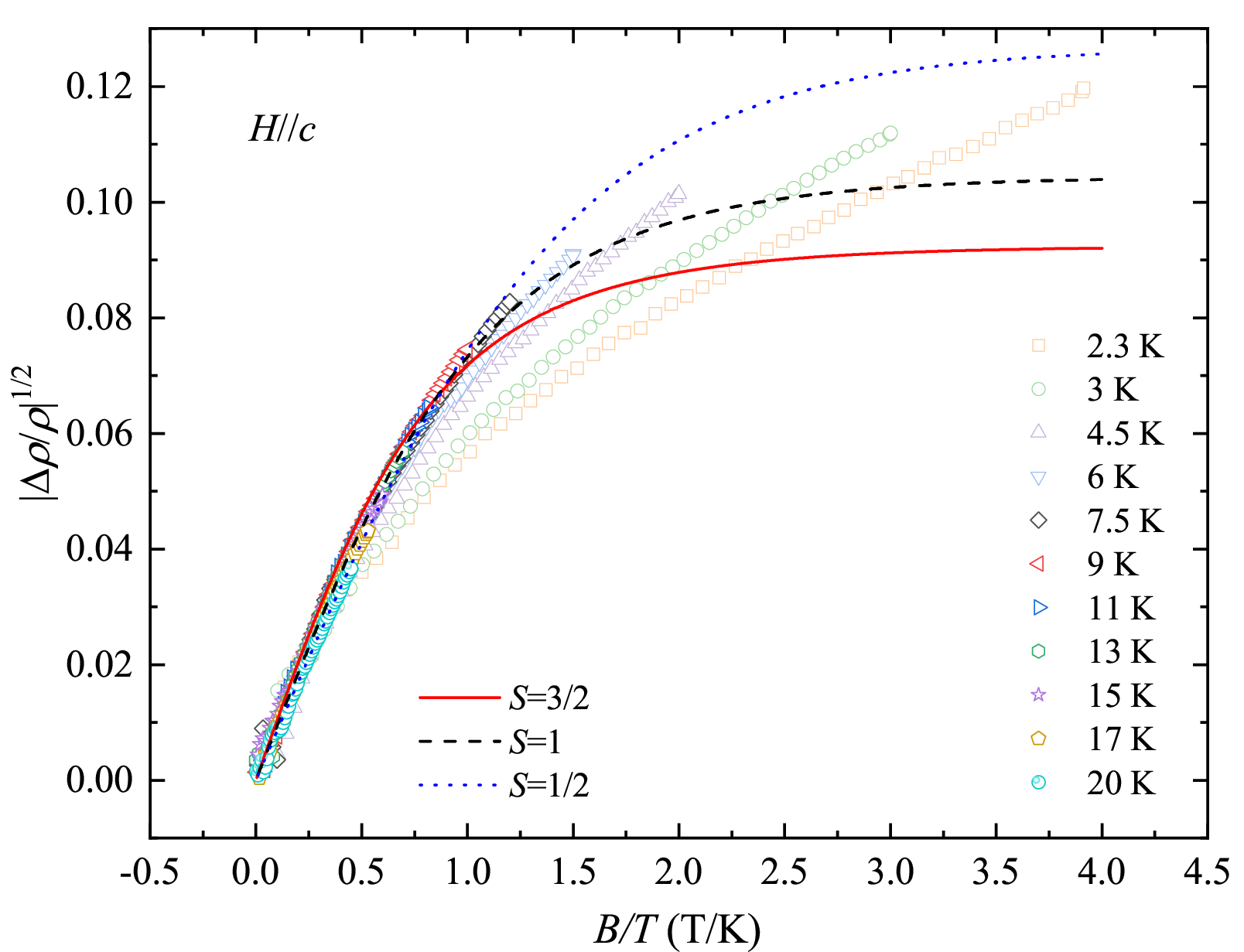}
\caption{The square root of $|\Delta \rho/ \rho_{0}|$ versus $B/T$ at low temperatures. The solid, dashed and dotted lines in red, bleak, and blue represent the Brillouin function fits (eq.~\ref{eKondoMR1}) for $T\geqslant7.5$~K with $S=3/2$, 1, and 1/2, respectively.}
\label{Figure7}
\end{figure}
In a Kondo system, for $T\geqslant T_K$, the leading term of the magnetoresistance can be given by the square of the magnetization $M^2$ to both second~\cite{yosida1957anomalous} and third order~\cite{beal1968negative} perturbation expansions of the $s$-$d$ exchange Hamiltonian. Considering that the magnetization can be described by the Brillouin function, we have
%for $T>T_{K}$, the square root of the absolute value of MR can be expressed as \cite{beal1968negative}

% Eq.4
%
\begin{equation}
\begin{aligned}
\bigg|\frac{\Delta \rho}{\rho_{0}} \bigg|^{1/2} =\lambda B_S(\frac{gS\mu_{B}B}{k_{B}T}).
\label{eKondoMR1}
\end{aligned}
\end{equation}
Where $\lambda$ is a constant, $B_S(x)=\frac{2S+1}{2S}coth(\frac{2S+1}{2S}x)-\frac{1}{2S}coth(\frac{1}{2S}x)$ is the Brillouin function, $B=\mu_0H$ is the magnetic flux density, $S$ is the magnetic impurity spin, $g=2$ is the Lander factor, $\mu_{B}$ is the Bohr magneton and $k_B$ is the Boltzmann constant. Here we assume that the orbital angular momentum is completely frozen since the influence of the crystalline field. It should be noted that the applicability of eq.~\ref{eKondoMR1} is limited to temperature $T\geqslant T_K$, since at temperatures below $T_K$, the magnetic moment of the impurity ions in the metal is gradually shielded and canceled out by the electron spin \cite{wilson1975renormalization}.

Fig.~\ref{Figure7} shows $|\Delta \rho/ \rho_{0}|^{1/2}$ as a function of $B/T$. The curves for $T\geqslant7.5$~K overlap each other. This means that the Kondo temperature $T_{K}$ is between 6$\sim$7.5~K, slightly smaller than the value obtained from the fits of the $\rho-T$ curve. The lines in Fig.~\ref{Figure7} represent the Brillouin function fits (eq.~\ref{eKondoMR1}) for $T\geqslant7.5$~K. The raw data for $T\geqslant7.5$~K lies between the fitting curves of the Brillouin function with $S=1$ and $S=1/2$, which means that $S$ is likely to be bewteen 0.5 and 1, slightly larger than that obtained from fits of the $\rho-T$ curve. The difference is probably mainly due to the large errors of the fits for the $\rho-T$ curve, since there are too many parameters. As metioned in the previous references~\cite{lu2024interlayer, taniguchi1995transport, qu2024bilayer, ouyang2024hund, cao2024flat, oh2023type}, the Ni $3d_{z^{2}}$ electrons are localized forming the localized magnetic moment, the Ni $3d_{x^{2}-y^{2}}$ electrons are itinerant. The interaction between electrons of these two orbitals may give rise to the Kondo effect. On the other hand, the fit of the magnetoresistance at low temperatures support that the spin of the magnetic impurities $S$ is close to 1. Since Ni is the only magnetic element in the \LNO\ thin film and its divalent ion has a spin $S=1$, there may be a small amount of $\rm{Ni^{2+}}$ impurities in the sample, causing the Kondo effect.

%For $T<T_{K}$, the parameter $H/T$ should be replaced by $H/(T+T_{K})$. The Hamann formula for zero-filed Kondo resistivity can be modified by the spin polarization ($1-B^{2}$) \cite{felsch1973magnetoresistivity}:

%%%%%%%%%%%%%%%%%%%%
% Figure 8
\begin{figure}[tb]
\includegraphics[width=\columnwidth]{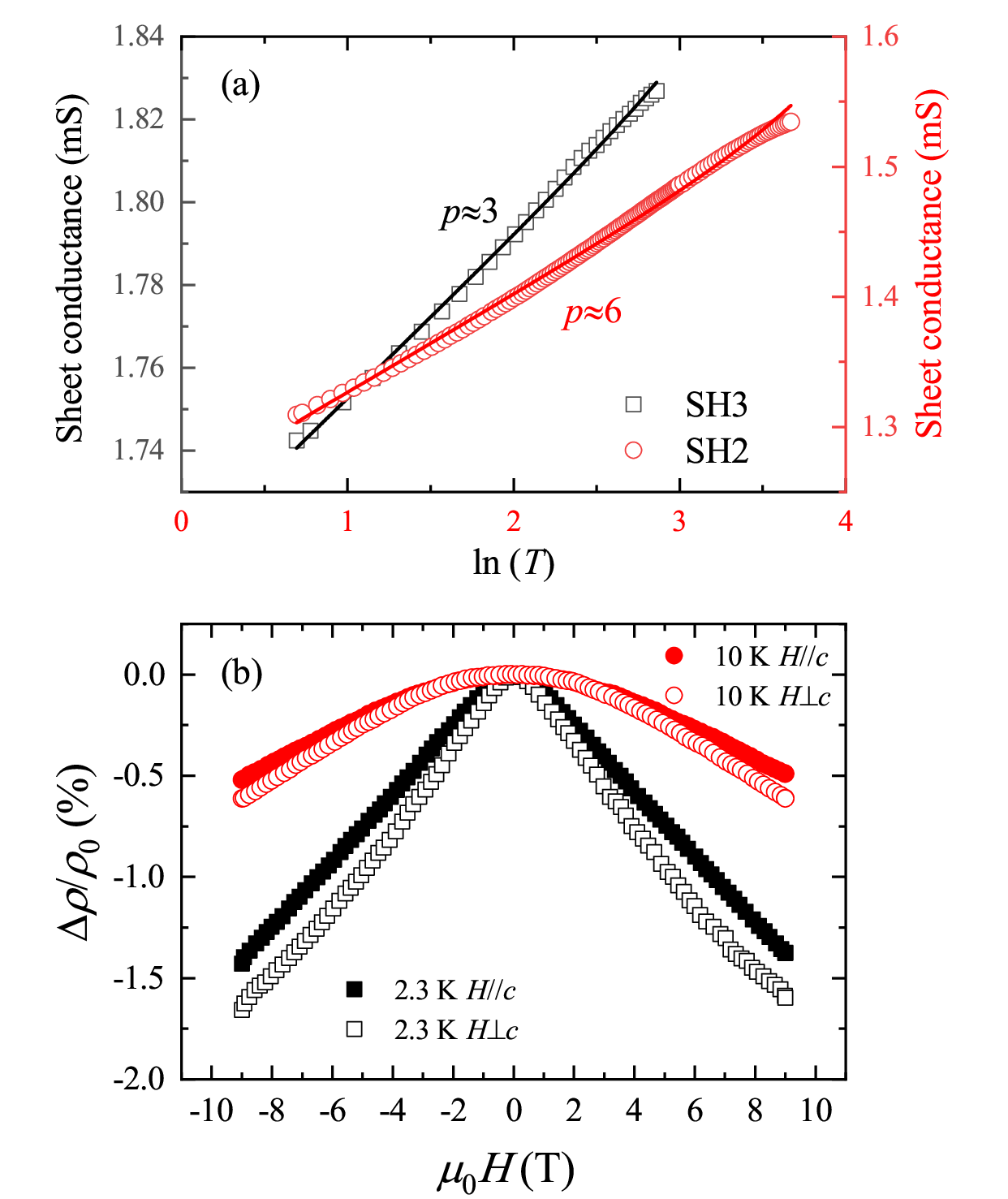}
\caption{(a) $\rho-T$ curves and fits to the model of weak localization (eq.~\ref{WLR}) at low temperatures for two metallic samples. (b) Magnetoresistance in the magnetic field applied parallel and perpendicular to the \LNO\ film $c$-axis for sample SH3.}
\label{Figure8}
\end{figure}

As mentioned above, Anderson weak localization effect may also cause a logarithmic increase of resistivity and negative MR at low temperatures. In the case for a 2D weak localization, we have~\cite{lee1985disordered}
% Eq.5
%
\begin{equation}
\begin{aligned}
\sigma_{2D}(T)=\sigma_{0}+a\frac{1}{T^2}+p\frac{e^2}{\pi h}ln(\frac{T}{T_0}).
\label{WLR}
\end{aligned}
\end{equation}
The inelastic scattering length is defined as $l_{in}\varpropto T^{-p/2}$, where $p$ depends on the scattering mechanism. For the major inelastic scattering due to electron-phonon scattering, we have $p=3$; for electron-electron scattering, we have $p=1$~\cite{lee1985disordered}. $\sigma_{0}+a/T^2$ is the Drude conductivity, where the $1/T^2$ term represents the contribution of Fermi liquid behavior. $T_0$ represents the lower length cutoff for diffusive motion and is related to the transport mean free path. Fig.~\ref{Figure8}(a) shows the sheet conductance versus ln($T$) of the metallic \LNO\ thin films at low temperatures, and the solid lines show the fits with eq.~\ref{WLR}. For sample SH3, we have $p\approx3$, but for sample SH2, $p\approx6$, which does not make sense. Considering that the mechanism that causes the resistivity upturn shouldn't be different for these two similar samples, we exclude the weak localization effect in the \LNO\ films.

Furthermore, weak localization effects generally appear in two-dimensional electronic gases or metallic thin films. The former is strictly a two-dimensional system, the latter can be regarded as two-dimensional when the film thickness is less than inelastic diffusion length. When a magnetic field perpendicular to the plane is applied, due to the change of electron relative phase, the intensity of weak localization decreases rapidly. Conversely, when the applied magnetic field ia parallel to the plane, the intensity of weak localization will not vary greatly. Unlike the weak localization effect, the magnetic impurities are randomly distributed in the crystal in Kondo system. Therefore, the variation of magnetoresistance in different magnetic field directions should be similar for the Kondo-like scattering. Fig.~\ref{Figure8}(b) shows the magnetoresistance for different magnetic field directions. In the two measured directions, the negative MR has similar variation tendencies, with only a slight difference in quantity. This also supports the Kondo-like explanation.

%%%%%%%%%%%%%%%%%%%%%%%%%%%%%%%%%%%%%%%%%%%%%%%%%%%%%%%%%%%%%%%%%%%%%%%%%%%%%%
%
% Conclusions
%
\section{SUMMARY}
In summary, we have successfully synthesized (00$l$)-oriented \LNO\ thin films on \LAO\ single crystal substrates by PLD technique. We have obtained samples with weak insulating and metallic behavior by adjusting the laser energy and oxygen partial pressure during growth. The rocking curves show that the crystallinity of the metallic samples is significantly better than that of the weak insulating samples. Moreover, the $\rho-T$ curves of all the \LNO\ thin films in the temperature region of 2$\sim$300~K do not show any anomalies corresponding to the spin or the charge density waves that have been observed in bulks. This may be attributed to the influence of the compressive strain provided by the substrate. Hall resistance shows a linear field dependence with the dominant hole charge carriers, but the Hall coefficient exhibits strong temperature dependence. The MR above about 50~K is positive but very weak, indicating a weakened or absence of multiband effect. The Hall angle shows that the scattering rate is close to $1/\tau\propto T^{2.24}$, but the scattering rate derived from resistivity shows the temperature dependence with an exponent of about 1.5, which is explained by the involvement of correlation effect in the \LNO\ thin films. Moreover, the magnetoresistance at low temperatures is negative, showing the delocalization effect. Detailed analysis of the magnetoresistance suggests that the delocalization effect can be attributed to the Kondo-like scattering, rather than the Anderson weak localization. The Kondo-like scattering suggests a Hund's coupling between the  $d_{x^{2}-y^{2}}$ orbital with the $d_{z^{2}}$ orbital.

%%%%%%%%%%%%%%%%%%%%%%%%%%%%%%%%%%%%%%%%%%%%%%%%%%%%%%%%%%%%%%%%%%%%%%%%%%%%%%
%
% Acknowledgment
%

\begin{acknowledgments}
We thank B.~Yang and Y.~Z.~Tian for assistance with XRD and X-ray reflectivity analysis. Thank B.~Hao for help with the RSM measurement and analysis. We acknowledge financial support from the National Key Research and Development Program of China  (No. 2022YFA1403201), National Natural Science Foundation of China (Nos. 11927809, 12061131001). 
\end{acknowledgments}

%%%%%%%%%%%%%%%%%%%%%%%%%%%%%%%%%%%%%%%%%%%%%%%%%%%%%%%%%%%%%%%%%%%%%%%%%%%%%%%
% The bibliography (BibTeX)

\end{document}